\documentclass[a4paper,fleqn,usenatbib]{mnras}

\usepackage{newtxtext,newtxmath}

\usepackage[T1]{fontenc}
\usepackage{ae,aecompl}

\usepackage{graphicx}	
\usepackage{amsmath}	
\usepackage{amssymb}	

\usepackage{verbatim}	

\newcommand{\Hi}{H\,{\sc i}}

\newcommand{\Di}{D\,{\sc i}}
\newcommand{\DtH}{\Di/\Hi{}}
\newcommand{\DHp}{$\left(\mathrm{D/H}\right)_\mathrm{p}$} 					
\newcommand{\Cii}{C\,{\sc ii}}

\newcommand{\Oi}{O\,{\sc i}}

\newcommand{\Siii}{Si\,{\sc ii}}
\newcommand{\Siiv}{Si\,{\sc iv}}
\newcommand{\siiv}{Si\,{\sc iv}}
\newcommand{\Feiii}{Fe\,{\sc iii}}
\newcommand{\Feii}{Fe\,{\sc ii}}

\newcommand{\omegab}{$\Omega_bh^2$} 
\newcommand{\kms}{\, \mathrm{km\, s}^{-1} }
\newcommand{\sqcm}{\, \mathrm{cm}^{-2} } 					
\newcommand{\cm}{\, \mathrm{cm} } 					
\newcommand{\ang}{\, \AA{}}

\newcommand{\chidof}{$\chi^2/\mathrm{dof}$}

\newcommand{\vpfit}{\texttt{VPFIT}}

\newcommand{\cloudy}{\texttt{Cloudy}} 
 
\newcommand{\Planck}{the {\sl Planck Surveyor}} 

\newcommand\eqnref[1]{%
 Eqn.~\ref{eqn:#1}}

\newcommand\tabref[1]{%
Tab.~\ref{tab:#1}}

\newcommand\figref[1]{%
Fig.~\ref{fig:#1}}

\newcommand\secref[1]{%
Sec.~\ref{sec:#1}}

\newcommand\appref[1]{%
App.~\ref{app:#1}}

\usepackage[usenames]{xcolor}

\title[Precise deuterium measurement at z=3.572]{A precise deuterium abundance: Re-measurement of the $z=3.572$ absorption system towards the quasar PKS1937--101}

\author[Riemer-S\o{}rensen et al.]{S. Riemer-S\o{}rensen$^{1,2}$\thanks{Email: signe.riemer-sorensen@astro.uio.no}, 
S. Kotu\v{s}$^{3}$, J. K. Webb$^{4}$, K. Ali$^{5},$
 \newauthor
V. Dumont$^{4,6}$, M. T. Murphy$^{3}$, R. F. Carswell$^{7}$\\
$^{1}$Institute of Theoretical Astrophysics, The University of Oslo, Boks 1072 Blindern, NO-0316 Oslo, Norway\\
$^{2}$ARC Centre of Excellence for All-sky Astrophysics (CAASTRO)\\
$^{3}$Centre for Astrophysics and Supercomputing, Swinburne University of Technology, P.O. Box 218, Hawthorn, VIC 3122, Australia\\
$^{4}$School of Physics, University of New South Wales, Sydney NSW 2052, Australia\\
$^{5}$International Centre for Radio Astronomy Research (ICRAR), University of Western Australia, 35 Stirling Hwy, Crawley, WA 6009, Australia\\
$^{6}$Department of Physics, University of California, Berkeley, California 94720-7300, USA\\
$^{7}$Institute of Astronomy, University of Cambridge, Madingley Road, Cambridge CB3 0HA, United Kingdom}

\date{Accepted XXX. Received YYY; in original form ZZZ}

\pubyear{2016}

\begin{document}
\label{firstpage}
\pagerange{\pageref{firstpage}--\pageref{lastpage}}
\maketitle

\begin{abstract}
The primordial deuterium abundance probes fundamental physics during the Big Bang Nucleosynthesis and can be used to infer cosmological parameters. Observationally, the abundance can be measured using absorbing clouds along the lines of sight to distant quasars. Observations of the quasar PKS1937--101 contain two absorbers for which the deuterium abundance has previously been determined. Here we focus on the higher redshift one at $z_\mathrm{abs} = 3.572$. We present new observations with significantly increased signal-to-noise ratio which enable a far more precise and robust measurement of the deuterium to hydrogen column density ratio, resulting in \DtH{} = $2.62\pm0.05\times10^{-5}$. This particular measurement is of interest because it is among the most precise assessments to date and it has been derived from the second lowest column-density absorber [$N($\Hi$)= 17.9\cm^{-2}$] that has so-far been utilised for deuterium abundance measurements. The majority of existing high-precision measurements were obtained from considerably higher column density systems [i.e.\ $N($\Hi$)>19.4\cm^{-2}$]. This bodes well for future observations as low column density systems are more common.
\end{abstract}



\section{Introduction}
Any presence of non-standard physics, e.g. from dark matter or additional neutrinos, during the epoch of Big Bang Nucleosynthesis (BBN) may change the conditions under which the light elements such as deuterium, helium and lithium formed and their resulting abundances \citep[e.g.][]{Steigman:2012, Nollett:2014, Boehm:2013,Archidiacono:2014}. While the observed abundances (apart from lithium) are in overall agreement with the Standard Model predictions, we need high-precision measurements \citep{Izotov:2014,Aver:2015}, to distinguish between more detailed scenarios \citep[e.g.][]{DiValentino:2013,Steigman:2013}. 

The abundance of deuterium traces the number density of baryons at early times, which can also be determined from the Cosmic Microwave Background (CMB) \citep[e.g.][]{Fixsen:2009}. The \Planck\ provides the highly precise measurement of $\Omega_bh^2 = 0.02225\pm0.00016$ (where $h$ is the dimensionless Hubble parameter $H_0 = 100h\kms\mathrm{Mpc}^{-1}$) at the time of recombination \citep[][]{PlanckXIII:2015}, which can be compared to the value obtained from BBN to infer any time evolution.

In this paper we present a new and precise measurement of the deuterium abundance in the absorption system at $z_{\rm abs}=3.572$ towards the quasar PKS1937--101 (B1950, emission redshift $z_\mathrm{em} = 3.787$ \citep{Lanzetta:1991}). The deuterium abundance has previously been determined in two well-separated absorption systems in the sight-line to PKS1937--101 at $z_\mathrm{abs} = 3.256$ \citep{Crighton:2004} and $z_\mathrm{abs} = 3.572$ \citep{Tytler:1996, Burles:1998a}. Here we focus on the $z_\mathrm{abs} = 3.572$ absorber, previously suggested as an ideal absorber for a \DtH{} measurement, due to its low metallicity, high column density, and simple velocity structure \citep{Tytler:1996}. The low redshift system was re-analysed in a companion paper \citep{Riemer-Sorensen:2015}. Since the first measurements were published, PKS1937--101 has been the target of extensive observations with both the Ultraviolet and Visual Echelle Spectrograph (UVES) at the Very Large Telescope (VLT) and the High Resolution Echelle Spectrometer (HIRES) at the Keck Telescope, so that the total exposure time has increased by almost a factor of 10.

The precision of the measurement presented here is similar to those of \citet{Cooke:2014}, but from a much lower column density absorber with $N($\Hi$)= 17.9\cm^{-2}$ versus $N($\Hi$) > 19.4\cm^{-2}$. 

In \secref{Observations} we present the observational data, and the analysis details in \secref{Analysis}. The results are presented in \secref{Results} with the details of the best fit model given in \appref{Model}. In \secref{Discussion} we discuss various caveats such as the probability of blending, and compare the new measurements to those from previous studies and also to the estimate of \omegab{} from the {\sl Planck} measurement of the CMB. We also investigate any cosmological implications before concluding in \secref{Conclusions}.

Unless otherwise stated, we quote uncertainties as 68 per cent confidence level, and column densities are quoted as $\log(N)$ where $N$ has the unit of $\sqcm$.

\section{Observations} \label{sec:Observations}
The observational data used in this paper are listed in \tabref{data}. It include all the observations listed in Tab.~1 of \citet{Riemer-Sorensen:2015} plus an additional publicly available Keck observation\footnote{Keck Observatory Archive \url{http://www2.keck.hawaii.edu/koa/public/koa.php}}. The observations are reduced and continuum-fitted using standard procedures as described in \citet{Riemer-Sorensen:2015}. In section \secref{continuum} we discuss possible continuum-level uncertainties and how we account for them.

The individual quasar exposures taken with similar grating settings and slit-widths (i.e.\ resolving powers) were combined with inverse-variance weighting. This provided a total of five final spectra which, if combined, would completely cover the wavelength range $4100-6400$\ang. The average signal-to-noise ratio of the fitted Lyman regions varies from 14 to 65 for pixel sizes of $2.5-4.3\kms$ (the individual values are given in \tabref{sn}).

\begin{table*}
\begin{minipage}{158mm}
\caption{Observations included in the analysis (including those of Tab. 1 in \citet{Riemer-Sorensen:2015})}\label{tab:data} 
{\centering
\begin{tabular}{@{}l l l l l l l@{}}
\hline
Date	& Primary 		& Instrument		& Settings	\footnotemark[1]		& Resolving 	& Resolution\footnotemark[2] 	& Observation \\
	& investigator	& 		& 				& power 	& $\sigma_v [\kms]$		& time [ks]			 \\ \hline
1996-08-09	& Songaila		& Keck LRIS		& w=0.7\arcsec and 1.5\arcsec		& 1500, 300	& 400 			& 2.4, 2.7	\\
1997-10-02	& Cowie		& Keck HIRES	& C5 (1.148\arcsec, 4000/6480\ang)		& 37000	& 3.5			& 4x2.4\\ 
1997-10-03	& Cowie		& Keck HIRES	& C5 (1.148\arcsec, 3910/6360\ang)		& 37000	& 3.5			& 4x2.4\\ 
1997-10-04	& Cowie		& Keck HIRES	& C5 (1.148\arcsec, 3910/6360\ang)		& 37000	& 3.5			& 2x2.4 + 1x1.4\\ 
2005-07-01	& Crighton		& Keck HIRES	& B5 (0.861\arcsec, 3630/8090\ang)		& 49000	& 2.8			& 6x3.6 \\ 
2005-08-12	& Tytler		& Keck HIRES	& C5 (1.148\arcsec, 3790/6730\ang)		& 37000	& 3.5			& 2x6.4 + 1x6.0\\ 
2006-04-10	& Carswell\footnotemark[3] & VLT UVES	& DICHR\#1 (1.0\arcsec, 3900/5800\ang)	& 45000	& 2.8 			& 1x5.4 \\
2006-06-01	& Carswell\footnotemark[3] & VLT UVES	& DICHR\#1 (1.0\arcsec, 3900/5800\ang)	& 45000	& 2.8 			& 2x5.4 \\
2006-06-25	& Carswell\footnotemark[3] & VLT UVES	& DICHR\#1 (1.0\arcsec, 3900/5800\ang)	& 45000	& 2.8 			& 1x5.4 \\
2006-07-21	& Carswell\footnotemark[3] & VLT UVES	& DICHR\#1 (1.0\arcsec, 3900/5800\ang)	& 45000	& 2.8 			& 1x5.4 \\
\hline
\end{tabular}}

\footnotemark[1]{Slit width for LRIS, slit width, cross-disperser angle and central wavelength for HIRES, and slit width and central wavelength of the blue/red arms for UVES}\\
\footnotemark[2]{1-$\sigma$ velocity width of the resolution element, $\sigma_v$, as determined by illuminating the instrument with a Thorium-Argon lamp. The individual $\sigma_v$ have been $\chi^2$ optimised as described in \secref{resolution}.}\\
\end{minipage}
\end{table*}

\begin{table}
\caption{Signal-to-noise ratios of the fitted Lyman series regions }\label{tab:sn}
\begin{tabular}{@{}l r r r r r r}
\hline
Transition		& Width  	& HIRES	& pixel size 	& UVES	& pixel size \\
		& [$\kms$] 	& S/N	& [$\kms$] 	& S/N	& [$\kms$] \\ \hline
Lyman $\alpha$	& 1104	& 60	& 3.2	& 46	& 2.5 \\
Lyman $\beta$	& 563	& 65	& 3.9	&  --- 	& ---\\
Lyman $\gamma$	& 539	& 36	& 4.1	& 21	& 2.5 \\
Lyman $4$		& 450	& 23	& 4.1	& 14 	& 2.5\\
Lyman $5$		& 406	& 36	& 4.2	& 27 	& 2.5\\
Lyman $6$		& 366	& 36	& 4.2	& 26 	& 2.5\\
Lyman $7$		& 391	& 28	& 4.3	& 23 	& 2.5\\
Lyman $8$		& 614	& 40	& 4.3	& 28 	& 2.5\\
Lyman $9$		& 278	& 29	& 4.3	& 21 	& 2.5\\
\hline
\end{tabular}
\end{table}

\section{Analysis and results} \label{sec:Analysis}
A more complete description of the analysis methods is given in \citet{Riemer-Sorensen:2015}. Here we provide additional details specific to the analysis described in this paper.

\subsection{Spectral fitting}
\begin{figure*}
\centering
\includegraphics[width=0.99\textwidth,trim={0.1cm 0.2cm 0.7cm 0.3cm},clip]{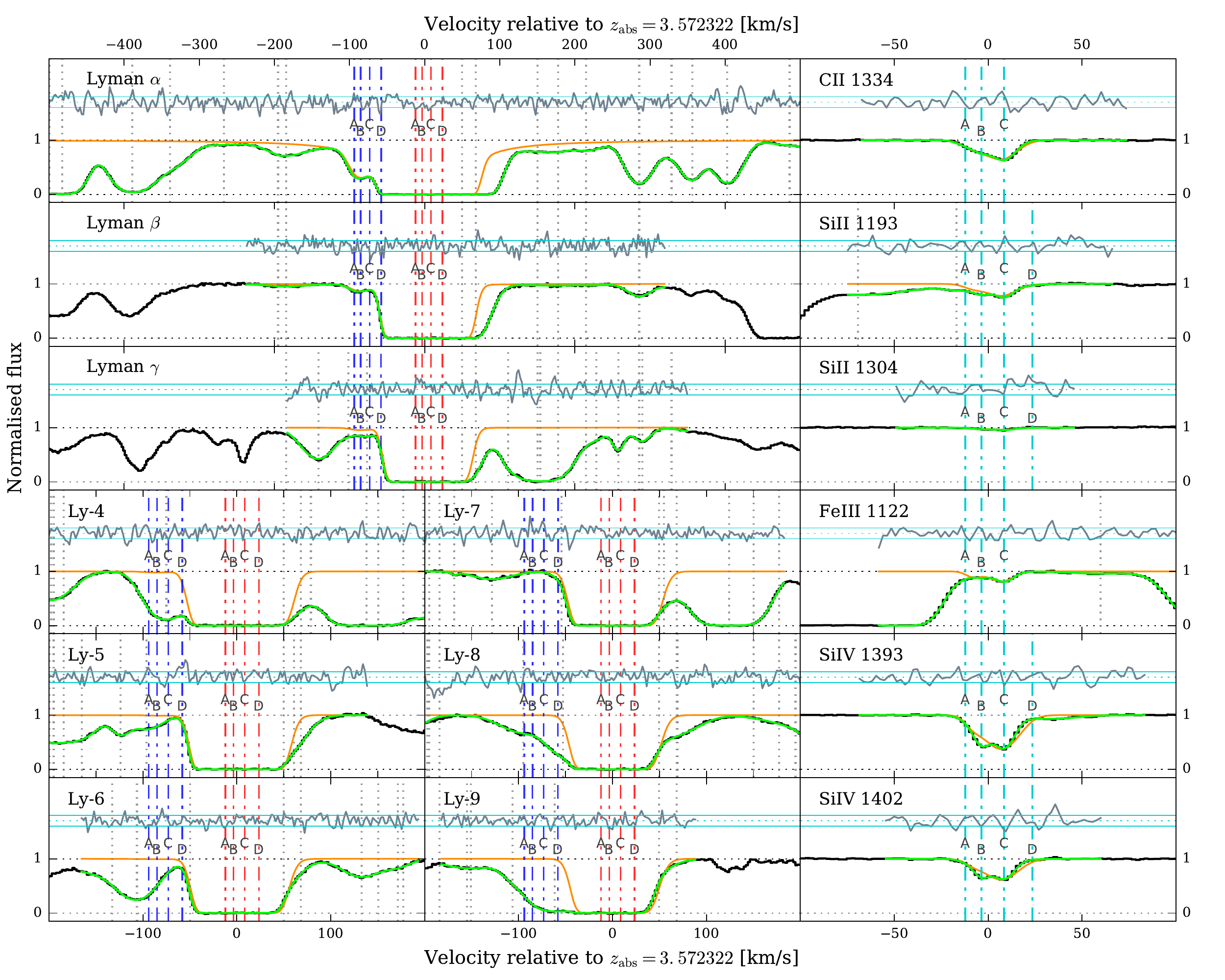}
\caption{The composite spectrum, created by combining the four different exposure stacks used in the analysis with variance weighting (thick black), the final model fit (thick green/light grey) and the residuals between the two (grey, above the spectrum in each panel) normalised by the 1-$\sigma$ error array (indicated by light blue horizontal lines). This composite spectrum is not used in the analysis; it is for visualisation purposes only and the analysis uses the four exposure stacks simultaneously to constrain the model fit parameters. The vertical dot-dash lines mark the velocity components A--D: red/light grey for \Hi, blue/dark grey for \Di, light blue/light grey for heavy elements. Interloping \Hi{} absorption is marked by light grey vertical dotted lines. The model from \citet{Tytler:1996} (without blends) is over-plotted (thin orange/grey). It is clear that we find a different structure particularly for the metals.}
\label{fig:spectrum}
\end{figure*}

The five individual spectra (\secref{Observations}) are fitted simultaneously but separately, rather than being combined into a single, composite spectrum as is more commonly the approach. For visualisation we use a variance-weighted stacked spectrum of the HIRES and UVES spectra as shown in \figref{spectrum}. We fit six heavy element transitions: \Feiii{} 1122, \Siii{} 1193, \Siii{} 1304, \Siiv{} 1393, \Siiv{} 1402, and \Cii{} 1334, and nine Lyman transitions (from Lyman $\alpha$ to Lyman 9) together with the saturated Lyman limit in the lower resolution spectrum from the Low Resolution Imaging Spectrometer (LRIS) at the Keck Telescope. We include the LRIS Lyman limit data to better constrain the hydrogen column density.

To optimise the accuracy of our measurement we solve for the total column density in the absorption complex rather than the column densities of the individual subcomponents. We force the column density ratio of \Di/\Hi{} to be identical across all subcomponents, which corresponds to assuming that any deuterium depletion mechanisms act uniformly on all components. This is a reasonable assumption as the total metallicity of the absorber is less than 1/100th of the solar value (\secref{metallicity}). The column densities of the other species are allowed to vary between components.

To assess whether the scatter in current deuterium measurements (see \tabref{measurements} and \figref{correlations}) might be correlated with the observing telescopes, we also fitted the model to the spectra from Keck alone and from VLT alone, in addition to the entire data set. 

\subsection{Velocity shifts and contiuum}\label{sec:continuum}
As in \citet{Riemer-Sorensen:2015}, we explicitly allow for small velocity shifts between individual exposures which are expected to arise from slit-centering and wavelength calibration differences. We effectively force all regions from a given spectrum to have identical shifts. The only exception is the VLT spectrum of the Lyman 5 region where the preferred velocity shift ($\approx 1.1\kms$) differs to the other regions in the same spectrum ($\approx 0.5\kms$) and we have introduced that shift as an extra free parameter during the $\chi^2$ minimisation process. Allowing independent shifts in any of the other regions did not improve the $\chi^2$ per degree of freedom ($\chi^2$/dof).

To account for possible uncertainties in the continuum level determination, we fit a local continuum in each region. Where a reasonable amount of unabsorbed continuum is available on both sides of the line, we allow for a local slope on the continuum (\Cii{} 1334 and Lyman $\beta$, 5, 7, 8), but otherwise we keep the slope fixed at zero (\Feiii{}, \Siii, \Siiv{} and Lyman $\alpha$, $\gamma$, 4, 6, 9).

Both the continua and the best fit velocity shifts are given in \tabref{model}, and we apply them before combining the spectra for visualisation as shown in \figref{spectrum}.

\subsection{Spectral resolution} \label{sec:resolution}
The spectral resolution in velocity units, $\sigma_v$, is given in the telescope manuals\footnote{\url{http://www2.keck.hawaii.edu/inst/hires/slitres.html}, \url{http://www.eso.org/sci/facilities/paranal/instruments/uves/inst.html} for the individual settings based on} based on exposures with a Thorium-Argon lamp. For science exposures of the quasar, the slit may not be uniformly filled, unlike for the Thorium-Argon exposures. This may lead to a difference between the Thorium-Argon line width and the actual spectral resolution. We explored this possibility by varying the spectral resolution of each individual fitting region so as to minimise the overall $\chi^2$ for the fit while keeping all other parameter values fixed. 
Of the 72 individually fitted regions, the resolution was changed from the Thorium-Argon estimate for 14, of which only three required more than a 20 per cent adjustment. After this optimisation the resolutions were kept fixed during the fit. Re-running the fit after the adjustments of the resolution improved the total \chidof{} by approximately 0.15.

\subsection{Models} \label{sec:models}
The absorption signature just bluewards of the strongest hydrogen lines (Lyman $\alpha$, $\beta$ and $\gamma$) is visible in the composite spectrum in \figref{spectrum}. Its velocity is offset by $-88\pm13\kms$ from the main absorption, consistent with the expected deuterium--hydrogen separation of $82\kms$. Further, this absorption line is significantly narrower than other nearby Lyman lines, but significantly broader than is typical for heavy element absorption lines. We thus interpret this line as being due to \Di{} (see also \secref{blends}).

The composite spectrum in \figref{spectrum} shows that the Si{\sc \,iv} absorption comprises at least two velocity components, which are both slightly asymmetric. We model this with four Voigt profiles (velocity components A, B, C, D) which we require to have identical redshifts, temperatures, and turbulent $b$-parameters across all species, but individual summed column densities. Only the column density ratio of \Di{} to \Hi{} is assumed to be the same for components A--D, the remaining relative abundances are free to vary. The B and C components are clearly visible in all heavy elements, while A and D have lower column densities and D is not required to adequately fit the weaker metal transitions of \Cii{} and \Feiii{} for which the column densities drop below the threshold value of $\log(N)<8$ and are removed from the fit.

Adding additional components to the system did not improve the \chidof{} significantly. To avoid biasing, the final model for the absorption system was selected based on \chidof{} without checking the impact on the D/H ratio.

The presence of multiple heavy element species makes it possible to simultaneously fit for both temperature and $b$--parameters for the main components (A to D). 

The best fit \chidof{} is $1.04$ with the resulting parameters given in \tabref{model} and the \Hi{} and \Di{} column densities given in \tabref{results}.

The species of \Cii{}, \Siii{}, \Feiii{} and \Hi{} have significantly lower ionisation potentials than \Siiv{}, which therefore may not always trace the full velocity structure of the lower ionisation species \citep{Wolfe:2000,Fox:2007}. In locations where we find \siiv{} there is also bound to be hydrogen, but it may not be in the form of \Hi. In the model fitting this is taken into account by keeping the column densities of the individual species unrelated. If the fit is good (based on \chidof) without \Hi{} for a given component in \Siiv{} \vpfit{} will automatically remove the component. Consequently, the probability of finding the correct velocity structure is higher when we use the entire range of metals available, and including the higher ionisation species should not bias \Di/\Hi{}. We note that all components present in \Siiv{} are also present in \Siii, but with the velocity structure much better resolved in \siiv.
Refitting without \Siiv{} leads to a simpler velocity structure with only two components, but also a significant increase in \chidof{} e.g. 1.19 to 1.59 for the \Cii{} regions. We conclude from this that \Siiv{} helps significantly in determining the velocity structure. 

\Oi{} is not clearly present in the individual spectra, but when fitting all spectra simultaneously, we obtain a non-zero summed column density based on \Oi{} 1302 under the assumption of constant \Oi/\Hi{} across all components (see \tabref{model}). The \Oi{} column density is used as input for \cloudy{} simulations in \secref{metallicity} to determine the metallicity of the absorber.

 \subsection{Fitting the Lyman limit}
\begin{figure*}
\centering
\includegraphics[width=0.99\textwidth]{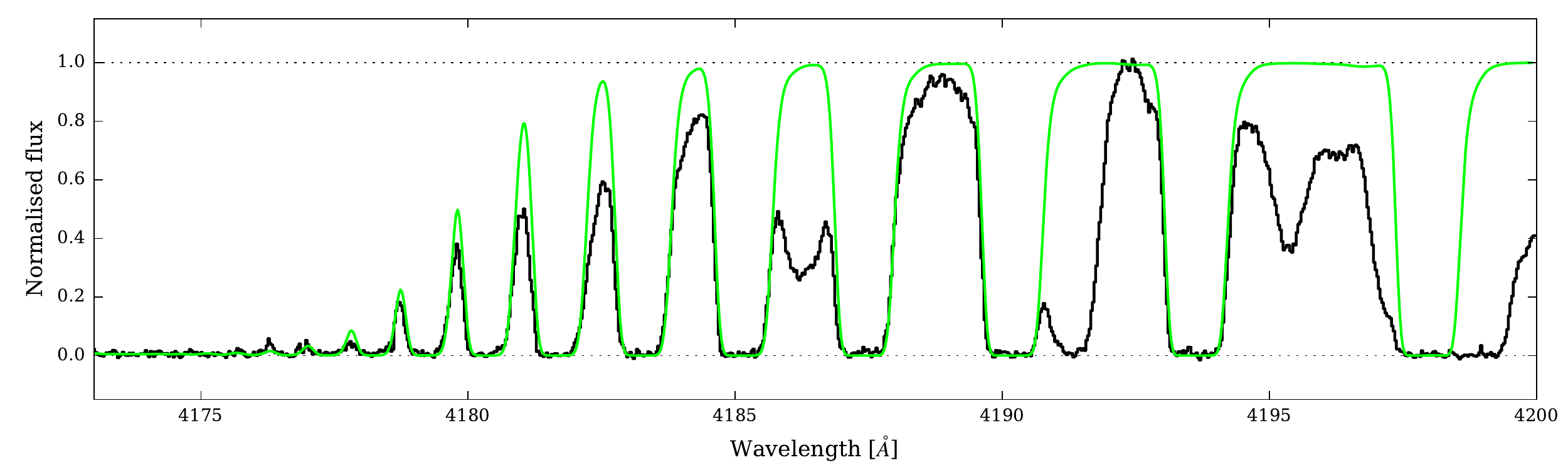}
\caption{The stacked high-resolution spectra (weighted by inverse variance) of the Lyman limit (thick black) with the best fit model (thin green/grey) based on the first nine Lyman transitions. It is clear that the model and the limit are consistent, but many blends are needed to fully model the limit and the continuum is not well defined.}
\label{fig:limit}
\end{figure*}

The high-resolution spectra of the Lyman limit contains many blended Lyman series lines which makes it harder to establish a reliable continuum level and complicates the modelling. Instead we verify that the model derived using the first nine lines in the Lyman series is consistent with the observed Lyman limit data as shown in \figref{limit}. 

\subsection{Resulting deuterium and hydrogen column densities}\label{sec:Results}
\tabref{results} gives the resulting column densities of \Di{} and \Hi{} and the ratios derived using the different data subsets. The main result from combining these measurements is a \DtH{} column density ratio of $2.62\pm0.05\times10^{-5}$. 

The spectra from the two different telescopes separately provide consistent results within the 1-$\sigma$ uncertainties, with the Keck spectrum leading to a slightly higher estimate of the  \DtH{} ratio: (\DtH)$_\mathrm{Keck} = 2.70\pm0.16\times10^{-5}$ compared to(\DtH)$_\mathrm{VLT} = 2.58\pm0.18\times10^{-5}$. This differences is much less than the combined uncertainties involved and there is no evidence to suspect any systematic deviation between the two telescopes.

The uncertainties quoted above and in \tabref{results} are only statistical; they derive from the signal-to-noise ratio of the spectra and do not include any systematic error estimates. However, the larger values of $\chi^2/$dof for the alternative models in \secref{models}, and the agreement between the results from different datasets, both suggest that the overall measurement is robust.

\begin{table*}
\begin{minipage}{168mm}
\caption{Best fit parameters of the models described in \secref{models}} \label{tab:results}
{\centering
\begin{tabular}{@{} l c c c c c@{}}
\hline
Spectra		& \chidof{}		& $\log(N$(\Di))				& $\log(N$(\Hi))					& $\log(N(\mathrm{D})/N(\mathrm{H}))$	& \DtH{} [$\times 10^{-5}$] 		\\ \hline
{\bf All}		& {\bf 1.04}	& {$\bf 13.344 \pm0.0056$} 	& $\boldsymbol{17.925\pm0.0063}$	& $\boldsymbol{-4.581\pm0.008}$	& $\boldsymbol{2.62\pm0.051}$	\\
Keck+LRIS	& 1.00 		& $13.357	\pm0.0244$		& $17.925\pm0.0066$			& $-4.568\pm0.025$		& $2.70\pm0.157$				\\
VLT+LRIS	& 1.12 		& $13.332	\pm0.0287$		& $17.921\pm0.0068$			& $-4.589\pm0.030$		& $2.58\pm0.175$				\\
\hline
\end{tabular}}

Note that we present additional significant figures for the parameter estimates in this table only to allow reproduction of our final uncertainty estimates in the \DtH{} in the final column.
\end{minipage}
\end{table*}

\subsection{Metallicity} \label{sec:metallicity}
Following the method from \citet{Riemer-Sorensen:2015} we use \cloudy{} simulations to determine the hydrogen number density and the metallicity of the main absorbers \citep{Ferland:2013}. In the \cloudy{} modelling we used a plane parallel geometry and a Haardt-Madau HM05 model\footnote{Described in Sec. 6.11.8 of the \cloudy{} documentation \url{http://www.nublado.org/browser/ branches/c13_branch/docs/hazy1_c13.pdf}.} as the ultraviolet background source. We generate a grid of models with the hydrogen number density bounded by $-5 < \log(n_\mathrm{H}[\mathrm{cm}^{-3}]) < 2.04$, where the high density limit comes from the upper limit to \Cii{}*/\Cii{} ratio determined as in \citet{Riemer-Sorensen:2015}.

\begin{figure*}
\centering
\includegraphics[width=0.99\textwidth]{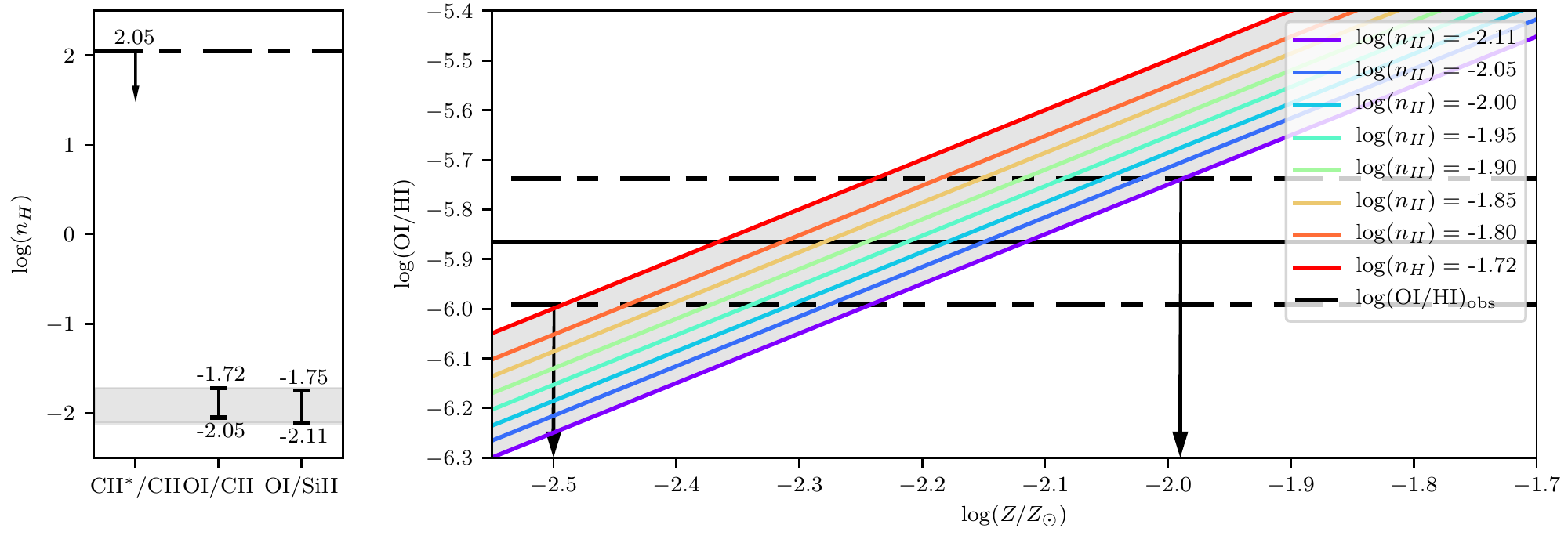}
\caption{{\bf Left:} The particle density range, $\log(n_\mathrm{H})$, inferred using \cloudy{} photoionization models based on the column density ratios derived from the quasar spectra using \vpfit{}. The upper limit from \Cii*/\Cii{}$=0.16$ assumes fully neutral hydrogen \citep[Eqn. 61 in][]{Bahcall:1968}. In practice, partial ionization is possible, which will decrease the actual upper limit, in agreement with the values obtained from \Oi/\Cii{} and \Oi/\Siii{}. The grey shaded area indicate the limits on $\log(n_\mathrm{H})$ used in the subsequent analysis. {\bf Right:} \Oi/\Hi{} column density ratio as a function of metallicity for increasing particle densities (from top to bottom). The observationally derived \Oi/\Hi{} column density is given by the horizontal line and the dashed lines span the uncertainty. The arrows indicate the extreme metallicities that are compatible with the allowed range of $\log(n_\mathrm{H})$ for the observed column density.}
\label{fig:metallicity}
\end{figure*}

The \Oi/\Hi{} and \Siii/\Hi{} ratios are fairly insensitive to $n_\mathrm{H}$ and provide an allowable range of metallicity of $-3 \lesssim \log (Z/Z_\odot) \lesssim -1$ given the observational data. The left panel of \figref{metallicity} shows the $\log n_\mathrm{H}$ range obtained by comparing \cloudy{} models with the observed abundances. The lower ionisation species agree with one another, leading to a conservative number density constraint $-2.11 < \log(n_\mathrm{H}) < -1.72$.

From the Voigt profile fit we obtain a marginal detection of \Oi{} with a summed column density of $\log{N(\mathrm{\Oi})} = 12.060\pm0.127$. The \Oi/\Hi{} ratio is sensitive to the metallicity $Z$ within the given $n_H$ range as shown in the right part of \figref{metallicity} (arrows) and provides a constraint of $-2.50 < \log (Z/Z_\odot) < -1.99$.  The simplest of chemical evolution models, as investigated by \citet{Fields:2001}, indicates little to no depletion of primordial deuterium for $\log (Z/Z_\odot) < -1$, hence this system appears to be a good estimator of initial \Di/\Hi{} given the allowable metallicity range.

We check the consistency of the \cloudy{} models by comparing the output gas temperature with that estimated by \vpfit{}. The \cloudy{} models take into account the average properties of all four components and provide a temperature range of $T_{\cloudy{}}=14500-16300\,\mathrm{K}$. From \vpfit{} the temperature is best determined for the dominating individual components A and C. The resulting temperatures of $T_{\vpfit}^\mathrm{A}=16410\pm 370 \mathrm{K}$ and $T_{\vpfit}^\mathrm{C}=18650\pm658\mathrm{K}$, respectively, are based on absorption in all considered species. As can be seen, the temperature of models generated by \cloudy{} is in reasonable agreement with that estimated by \vpfit{}.

Integrating through the absorbing cloud, i.e. allowing for varying particle density through the absorbing cloud, and taking into account the metallicity uncertainty, the cloud size is estimated as $2.7\pm1.7\, \mathrm{kpc}$.

\newpage

\section{Discussion} \label{sec:Discussion}
\subsection{Hydrogen contamination of deuterium lines} \label{sec:blends}
It is possible that the line identified as deuterium may be contaminated by a weak hydrogen line. Assuming the hydrogen column density distribution to be a power law with index $\beta = 1.7$ \citep{Penton:2004} and the redshift distribution to be a single power law with index $\gamma=1.51\pm0.09$ and normalisation $\log(A)=0.72\pm0.08$ for $13.1<N($\Hi$)<14.0$ \citep{Kim:2013}, we can determine the probability of detecting a line within a given redshift and column density interval. We take the relevant redshift interval to be spanned by the $3\sigma$ edges of components A to D and D/H$=[10^{-5},10^{-4}]$, which for the determined hydrogen column density of $\log N($\Hi$)=17.925$ corresponds to $\log N($\Hi$)=[12.925,13.925]$ (\Hi{} and \Di{} have the same oscillator strengths for the presumed Lyman $\alpha{}$ blending feature). This leads to a conservative estimate of the blending probability of 3.7\%. If we consider only the dominant component (A), the probability reduces to 0.25\% supporting the claim that the observed feature is likely to be deuterium.

We also checked whether any heavy element lines from other absorbers along the line of sight fall in the regions fitted to obtain \DtH. We identified tentative systems with lines from commonly found heavy elements (Al, C, Fe, Mg, Si) at $z=[$0.56030, 0.56790, 0.60312, 0.89683, 3.00851, 3.09545, 3.25618, 3.29134, 3.45359, 3.55357, 3.57230]. The only potentially problematic blend is from \Feii{} 1144 at $z=3.095$ which falls very close to \Di{} in Lyman $\beta$. However, no other \Feii{} lines are present anywhere in the spectrum at this redshift, some of which have larger oscillator strengths than \Feii{} 1144. We therefore assume the \Feii{} 1144 at $z = 3.095$ is too weak to significantly bias the deuterium column density.

\subsection{Comparison with previous measurements} \label{sec:previous}
Using an independent sample of Keck HIRES exposures, and only two velocity components to model the hydrogen absorption, \citet{Burles:1998b} estimated the \DtH{} ratio in the same absorber to be \DtH=$3.3\pm0.3\times 10^{-5}$ without using the heavy element lines. The initial measurement presented in \citet{Tytler:1996} was improved with a \Hi{} column density measurement of $\log(N($\Hi$))=17.86\pm0.02$ based on the HIRES spectra with additional low-resolution spectra from LRIS and the Kast spectrograph on the Shane 3 meter Telescope at the Lick Observatory \citep{Burles:1997,Burles:1998b}. 
 
The discrepancies between our new measurement and that of \citet{Burles:1998b} may be due to several factors such as continuum placement, number of components, and assumptions about physical properties of the absorber.
 
The continuum placement has been discussed in the literature. \citet{Wampler:1996} suggested models with 3--6 times larger \Hi{} column densities while \citet{Songaila:1997} obtained $\log(N($\Hi$))<17.7$ from the LRIS spectrum used in this analysis. Without supplementary high-resolution spectra their continuum may have been poorly estimated. According to \citet{Burles:1997} the unabsorbed continuum was underestimated in \citet{Songaila:1997}, but comparing with the high-resolution spectra we find that it is more likely to be overestimated. However, as we treat the local continua as free parameters, the discrepancy with \citet{Burles:1998b} more likely origins in the different numbers of fitted components. We have over-plotted the initial model from \citet{Tytler:1996} on the stacked spectrum in \figref{spectrum} (without additional hydrogen blends). \citet{Burles:1998b} did not use any of the heavy element lines to derive \DtH{}, and consequently their ($\chi^2$-based) choice of two components relies purely on the Lyman lines and not all the accessible information. Instead, here we assume that \Siiv{} traces the \Hi{} velocity structure despite the difference in ionisation potentials. The assumption is based on visual similarities in the spectrum combined with a significant increase in \chidof{} if fitting a two-component model without \Siiv{} (e.g. 1.19 to 1.59 for the \Cii{} regions). Fitting a four-component model without \Siiv{} leads to very small column densities for two components and significantly increased uncertainties (see also details in \secref{models}).
 
\subsection{The deuterium sample} \label{sec:sample}
\tabref{measurements} provides an updated version of the deuterium sample given in Tab. 4 of \citet{Riemer-Sorensen:2015}, comparing the new \DtH{} measurement presented in this paper to the sample from \citet{Pettini:2008} combined with recent measurements with similar precisions from \citet{Fumagalli:2011}, \citet{Noterdaeme:2012}, \citet{Cooke:2014}, \citet{Riemer-Sorensen:2015} and \citet{Balashev:2016}. 

\begin{figure*}
\centering
\includegraphics[width=0.99\textwidth]{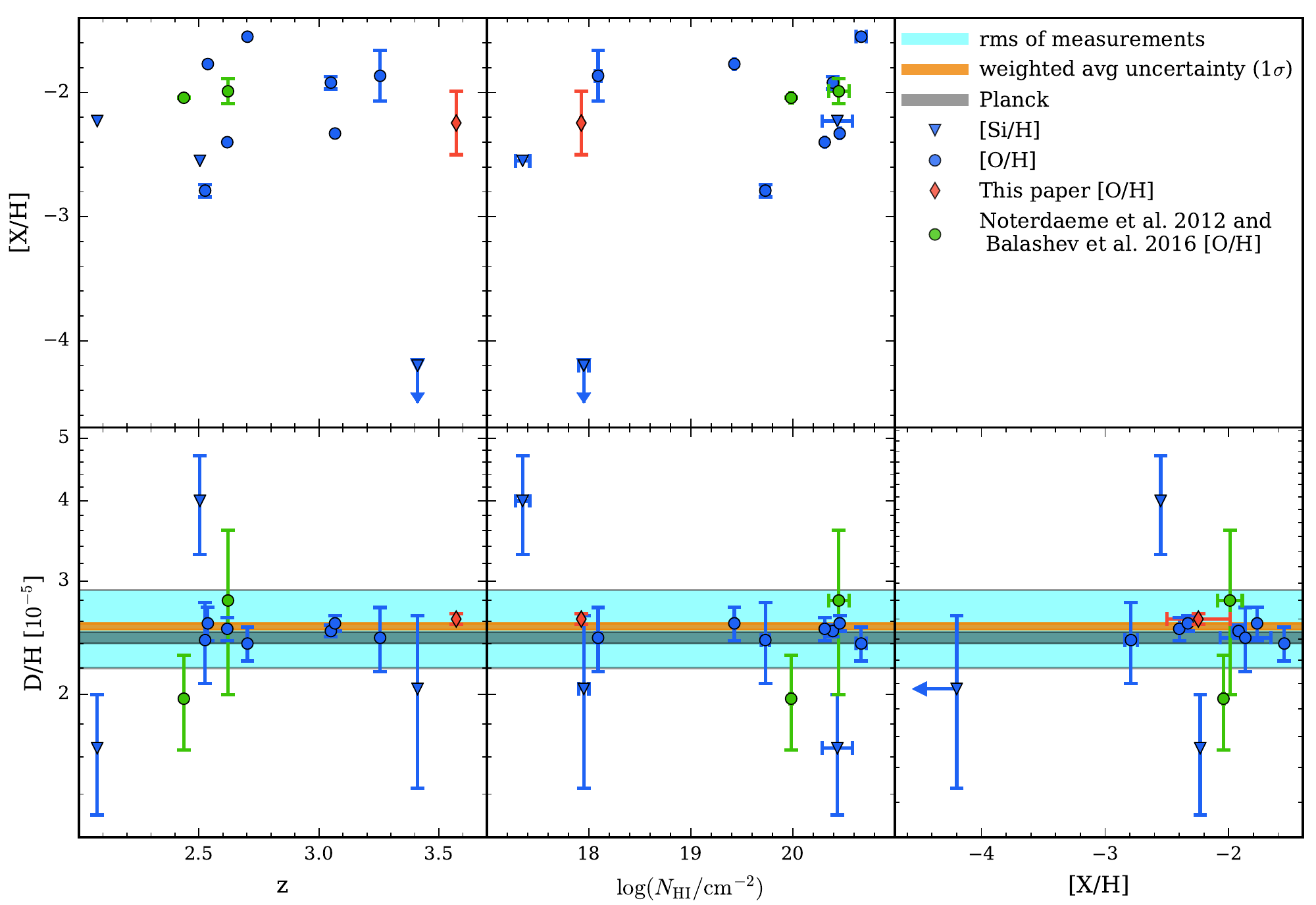}
\caption{The \DtH{} ratios and metallicities from the literature sample defined in \tabref{measurements} plotted as a function of redshift ($z$), \Hi{} column density ($\log(N($\Hi$/\cm^{-2}))$) and metallicity ([X/H]). The 1-$\sigma$ uncertainty in the weighted mean, \DtH$=2.55\pm0.03\times10^{-5}$, is shaded orange while the outer cyan shading indicates the root-mean-square deviation amongst the measurements. The inner dark shading indicates the constraint on \DtH{} derived from the \Planck{} measurements assuming standard BBN.}
\label{fig:correlations}
\end{figure*}

\figref{correlations} supersedes Fig. 4 from \citet{Riemer-Sorensen:2015}, illustrating the properties of the absorption systems for the results given in \tabref{measurements}. As in \citet{Riemer-Sorensen:2015}, we again find
no apparent correlation between any of the parameters, including deuterium and hydrogen column density versus redshift or metallicity.

\begin{table*}
\begin{minipage}{150mm}
{\centering
\caption{The sample of \DtH{} measurements considered robust in \citet{Pettini:2008} together with updated estimates in the same absorbers and more recent, similarly precise measurements from other absorbers.}\label{tab:measurements}
\begin{tabular}{@{}l c c c c c@{}}
\hline
Reference				& Absorption		& $\log (N($\Hi{}))             	& [X/H]            	& \DtH{}                   	& 100\omegab \\ 
				& redshift		&                                      	&                     	& $[\times 10^{-5}]$   	&                              \\ \hline
\citet{Burles:1998a}			& 2.504		& $17.4\pm0.07$            	& -2.55 Si       	& $4.00\pm0.70$       	& $1.66\pm0.18$    \\
\citet{Pettini:2001}			& 2.076		& $20.4\pm0.15$            	& -2.23 Si       	& $1.65\pm0.35$       	& $2.82\pm0.36$    \\
\citet{Kirkman:2003}			& 2.426		& $19.7\pm0.04$            	& -2.79 O       	& $2.43\pm0.35$       	& $2.24\pm0.20$    \\
\citet{Fumagalli:2011}			& 3.411		& $18.0\pm0.05$            	& -4.20 Si       	& $2.04\pm0.61$       	& $2.49\pm0.05$    \\
\citet{Noterdaeme:2012}			& 2.621		& $20.5\pm0.10$	& -1.99 O       	& $2.80\pm0.80$       	& $2.05\pm0.35$    \\	
\citet{Cooke:2014}, \citet{Pettini:2012}		& 3.050		& $20.392\pm0.003$	& -1.92 O       	& $2.51\pm0.05$       	& $2.19\pm0.02$    \\
\citet{Cooke:2014}, \citet{OMeara:2001} 	& 2.537		& $19.4\pm0.01$	& -1.77 O       	& $2.58\pm0.15$       	& $2.16\pm0.04$    \\
\citet{Cooke:2014}, \citet{Pettini:2008}		& 2.618		& $20.3\pm0.01$	& -2.40 O       	& $2.53\pm0.10$       	& $2.18\pm0.03$    \\
\citet{Cooke:2014}			& 3.067		& $20.5\pm0.01$	& -2.33 O       	& $2.58\pm0.07$       	& $2.16\pm0.03$    \\
\citet{Cooke:2014}, \citet{OMeara:2006} 	& 2.702		& $20.7\pm0.05$	& -1.55 O       	& $2.40\pm0.14$       	& $2.25\pm0.03$    \\
\citet{Riemer-Sorensen:2015}		& 3.255		& $18.1\pm0.03$	& -1.87 O		& $2.45\pm0.28$       	& $2.23\pm0.16$      \\	
\citet{Balashev:2016}			& 2.437		& $19.98\pm0.01$	& -2.04 O		& $1.97\pm0.33$	& $2.54\pm0.26$ \\
This work				& 3.572		& $17.925\pm0.006$ 	& -2.26 O		& $2.62\pm0.05$	& $2.14\pm0.03$    \\	
\hline
Weighted average\footnotemark[1]		& ---		& ---		& ---		& $2.55\pm0.03$	& $2.17\pm0.03$  \\
Unweighted average\footnotemark[1]		& ---		& ---		& ---		& $2.53\pm0.17$	& $2.18\pm0.08$ \\
\hline
\citet{PlanckXIII:2015}\			& ---		& ---		& --- 		& $2.45\pm0.05$	& $2.225 \pm 0.016$\\ \hline
\end{tabular}}

The conversion between \DtH{} and \omegab{} is based on nuclear rates from \citet{Coc:2015} for standard Big Bang Nucleosynthesis.
\footnotemark[1]{Without the \citet{Balashev:2016} and \citet{Noterdaeme:2012} measurements}
\end{minipage}
\end{table*}

The measurement from \citet{Balashev:2016} is derived from a fairly complicated absorption system under the assumption that the \Oi/\Hi{} ratio is identical for all components. The same assumption is applied in \citet{Noterdaeme:2012}. It is unclear whether this assumption is appropriate for a high precision measurement and we leave out these two measurements from further comparisons. 

A new sample of very precise deuterium abundance measurements ($\approx$4 per cent uncertainties) was presented by \citet{Cooke:2014}. They selected absorbers using narrow selection criteria to allow both precise and robust measurements; for example, they restricted the column density range to damped and sub-damped Lyman $\alpha$ systems, i.e.\ $\log (N$\Hi$)) \geq 19$. The obtained precision demonstrated the future prospects for deuterium as a cosmological probe. However, several important potential systematic errors remain, including the difficulties of modelling the uncertain velocity structure of individual absorbers, and estimating the uncertainties relating to continuum placement.

It is therefore necessary to obtain a significant sample of deuterium measurements so that it becomes possible to detect any plateau in D/H versus redshift or metallicity in order to obtain the primordial value, rather than relying on a small number of measurements where any intrinsic scatter cannot be reliably detected.

At $z=3.572$, our new measurement has the highest redshift in the sample and one of the lowest column densities. The precision we obtain is comparable to those of \citet{Cooke:2014} despite the new system being more complicated and lower column density. This demonstrates the future possibilities for a sample of high-resolution measurements as low column density absorbers are a lot more common than high column density absorbers. The last decade has seen a massive increase in the number and quality of high-resolution quasar spectra which have not yet been systematically searched for suitable low column density absorbers with visible deuterium lines. 

The weighted and unweighted averages centre on very nearby values of \DHp=$2.53\pm0.17$ and \DHp=$2.55\pm0.03$ respectively, which can be compared with the recent prediction from standard BBN of \DHp$=2.45\pm0.05$ \citep{PlanckXIII:2015,Coc:2015}. The deviation between the weighted average and \Planck{} value is $(2.55-2.45)/\sqrt{0.03^2+0.05^2}=1.7\sigma$ i.e. there is no significant difference. However the offset in the central values may indicate that some systematic error still remains despite the increasing data quality.

\figref{correlations} does not reveal any strong correlations between D/H and redshift, metallicity, or column density, but some scatter remains. According to recent chemical evolution modelling, we should expect some internal scatter in D/H measurements arising purely from the difference in the merger and star formation history of individual halos \citep{Dvorkin:2016}. Further high-quality measurements are needed to establish whether the different halo histories can explain the remaining scatter.

The most outlying D/H values in the sample are also the measurements with the highest uncertainties. If we define a high-precision sample by requiring $(\delta_{\mathrm{D/H}})/(\mathrm{D/H}) <0.15$, we automatically exclude the four most outlying points in \figref{correlations} and the weighted average become \DHp$=2.56\pm0.03$ consistent with the full sample value of $2.55\pm0.03$.

\subsection{The baryon fraction} \label{sec:baryon}
Assuming that the total deuterium-to-hydrogen ratio is reflected by \DtH{}, we can derive the primordial value \DHp{} because there are no sources of astrophysical production \citep{Epstein:1976, Prodanovic:2003} and the destruction rate in stars is low at the relevant redshifts and metallicities \citep{Romano:2006,Dvorkin:2016}.

\omegab{} can be obtained using fitting relations for standard BBN calculations \citep[e.g.][]{Simha:2008b,Steigman:2007,Steigman:2012,Coc:2015}. The most recent update with particular focus on the nuclear reactions gives \citep{Coc:2015}
\begin{equation}
10^5\mathrm{(D/H)} = (2.45\pm0.04)\left(\frac{\Omega_\mathrm{b}h^2}{0.02225}\right)^{-1.657} \, ,
\end{equation}
where the uncertainty of 0.04 reflects the measured uncertainties in the nuclear reaction rates. The results from the existing D/H measurements and the averaged values are listed in \tabref{measurements}. Several of the measurements in the table have percentage level statistical uncertainty, which is comparable to the uncertainty on the nuclear reaction rates. However, most of the quoted uncertainties do not take systematic uncertainties into account. In particular the uncertainties related to the number of modelled components and the risk of hydrogen blends at the position of deuterium requires high-resolution, high-quality spectra and are best controlled by having a large sample of high-precision measurements.

\subsection{Dipole fit} \label{sec:dipole}
Variations in fundamental constants such as the fine structure constant, the hadronic masses or binding energies would lead to variations in the light element abundances \citep{Dmitriev:2004,Flambaum:2007,Dent:2007,Berengut:2010}. \citet{Berengut:2011} investigated whether the observed scatter in the D/H measurements could be due to a dipole similar to the one observed for the fine structure constant \citep{Webb:2011, King:2012}. Although the significance of the fine structure constant dipole may be somewhat reduced by instrumental effects \citep{Whitmore:2015}, here we investigate whether a dipole in the same direction is preferred by the deuterium measurements and we also consider the dipole direction as a free parameter. The dipole is described by
\begin{equation} \label{eqn:dipole}
\log(N(D)/N(H)) = m_\mathrm{D}+d_\mathrm{D}r\cos (\psi(\phi_d,\theta_d) ) \, ,
\end{equation}
where $m_\mathrm{D}$ is the average observed deuterium abundance (the monopole), $d_\mathrm{D}$ is the magnitude of the dipole, $r=ct$ is the look-back distance, and $\psi(\phi_d,\theta_d)$ is the angle between a given observation with $(\mathrm{R.A., Dec.}) = (\phi, \theta)$ and the dipole direction ($\phi_d,\theta_d$) given by
\begin{equation}
\cos\psi = \cos(\phi-\phi_d)\cos(\theta)\cos(\theta_d)+\sin(\theta)\sin(\theta_d) \, .
\end{equation} 

\begin{figure}
\centering
\includegraphics[width=0.99\columnwidth]{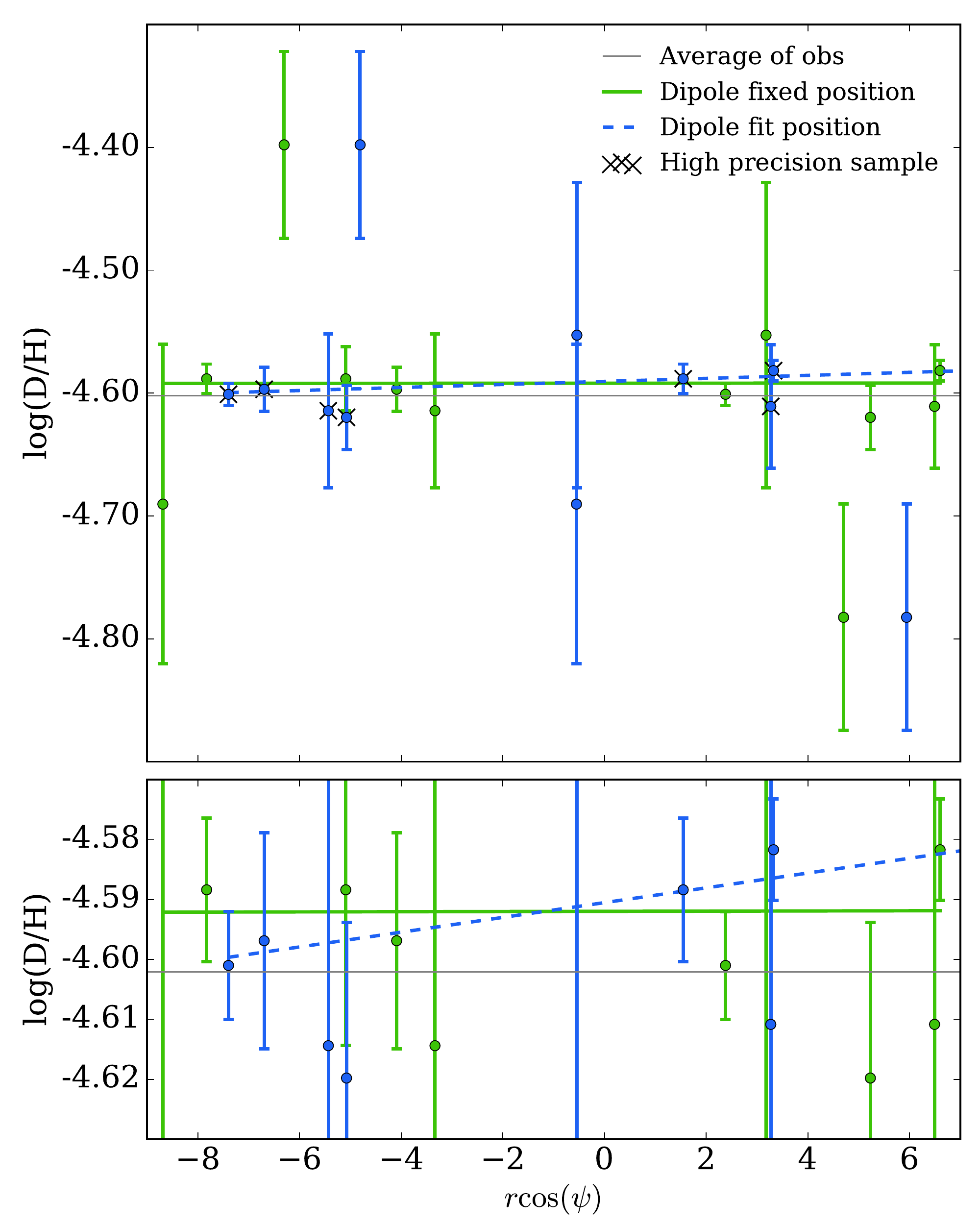}
\caption{The best fit to the dipole model of \citet{Berengut:2011} for fixed position (thick green line) and varying position (dashed blue line) as well as the observed abundances relative to the dipole directions (green and blue data points) and the average of the deuterium measurements (the monopole, thin grey line). The lower panel is a zoom of the y-axis. The crosses mark the high precision sample defined by less than 15\% uncertainty on D/H. The fits to the high-precision sample (not shown) lies very close to the full sample fits. The parameters are given in \tabref{dipole}.}
\label{fig:dipole}
\end{figure}

The best fits with and without fixing the direction of the dipole are shown in \figref{dipole} and the parameters given in \tabref{dipole}. The preferred slopes are close to zero with uncertainties larger than the preferred value and consequently consistent both with a small dipole and with no dipole. Despite the increase in sample size and precision, the scatter does not allow us to draw any firm conclusions about anisotropy.

We also fit to the high precision sample defined by less than 15\% uncertainty on D/H (\secref{baryon}). For fixed position this is consistent with no dipole. When fitting the high precision sample for the position, a dipole is preferred with around $2\sigma$ significance, but with a very large uncertainty on the direction and a $\chi^2<<1$ indicating too many free parameters in the fit.

\begin{table*}
\begin{minipage}{150mm}
{\centering
    \caption{\label{tab:dipole} Best fit parameters for fitting the variation in the deuterium measurements with the dipole model in \eqnref{dipole}}
    \begin{tabular}{l l l l l }
			& \multicolumn{2}{c}{Full sample} 			&\multicolumn{2}{c}{High precision sample}\\ 

			& Fixed position		& Fit position 	& Fixed position	& Fit position \\ \hline 
$m_\mathrm{D}$		& -4.592(6)			& -4.59(1) 		& -4.592(4)		& 4.590(6) \\
$d_\mathrm{D}$		& $0.000002(114)$		& 0.0012(14) 	& 0.00025(75) 	& 0.0014(6) \\
R.A. [h]			& $17.4\pm1.0$		& $24.0\pm5.3$ 	& $17.4\pm1.0$	& $0\pm2$ \\
Dec. [$\deg$]		& $ -61\pm10$		& $29.5\pm70.7$ 	& $ -61\pm10$	& $21.5\pm 32$\\
$\chi^2$			& 15.55			& 13.80		& 4.01		& 1.50\\
\chidof			& 1.56			& 1.72		& 0.67		& 0.38\\
$\Delta \mathrm{BIC}$\footnotemark[1]	& ---			& ---		& 2.1		& 2.1 \\
Evidence\footnotemark[2]		& ---			& ---		& Positive		& Positive \\
\end{tabular}}

\footnotemark[1]{Difference in Bayesian Information Criterion $\Delta\mathrm{BIC} = \mathrm{BIC}_\mathrm{slope}-\mathrm{BIC}_\mathrm{no \, slope}$.}\\
\footnotemark[2]{Evidence for preference of a zero slope model (no dipole) relative to a model with dipole based on the Bayesian Information Criterion.}
\end{minipage}
\end{table*}

\subsection{Inhomogeneities}
An interesting aspect of \DtH{} is that deuterium was produced when the universe was $100\sec$ old and the temperature around $10^9\,\mathrm{K}$. At that time the horizon size was $2ct = 6\times10^{12}\cm$. By today, this causally connected region has expanded by a factor of about $10^9/3$ and so spans $2\times10^{21}\cm$ or approximately 1 kpc. Regions larger than this (or collapsing down from regions larger than this) will have been causally disjoint at the time of deuterium synthesis. This implies that a sufficiently large sample of deuterium measurements will allow us to measure the homogeneity of the universe.

\subsection{Dark matter constraints}
Primordial element abundances may be very sensitive to the presence of the low-mass dark matter \citep{Steigman:2012, Nollett:2014, Boehm:2013,Archidiacono:2014,Stadnik:2015}. For example in scenarios where a scalar dark matter field interacts with the Standard Model fields during nucleosynthesis spatial variation in the \DtH{} ratio may arise \citep{Stadnik:2015}. Using the dipole constraints from \secref{dipole} we can improve the constraint on the product of the fractional energy density's spatial gradient with the interaction strength by a factor of roughly 40 relative to \citet{Stadnik:2015} as a consequence of the improved dipole constraint. The comparison of the calculated and measured deuterium abundances should lead to another breakthrough in the precision \citep{Berengut:2016}.

\section{Conclusions} \label{sec:Conclusions}
From the analysis of the $z_{\rm abs}=3.572$ absorption system in high-quality Keck and VLT spectra of quasar PKS1937--101, we make the following conclusions:
\begin{itemize}

\item  We find the \DtH\ ratio in this absorber to be $2.62\pm 0.05\times10^{-5}$ which corresponds to 100\omegab\ = 2.14$\pm$0.03 for standard BBN. This value deviates by $1.7\sigma$ from the Planck measurement of 100\omegab\ $= 2.225 \pm 0.016$ and is considered consistent. 

\item Independent fits to this absorption system using Keck and VLT spectra give consistent results.

\item The analysis presented here shows that lower column density systems can provide a precision on \DtH{} comparable to higher column density DLAs \citep{Cooke:2014}. This is important because the neutral hydrogen column density distribution in quasar absorption systems is a steep power law, with lower column density systems being more common. A statistically large sample of measurements is therefore feasible and necessary to reveal a plateau of primordial values as a function of, e.g., metallicity.

\item The spatial variation of the observed high-precision deuterium abundances is consistent with no dipole.
\end{itemize}

Deuterium abundance measurements using quasar absorption systems offer rare tests of the standard model of BBN and models with non-standard physics. While CMB measurements do offer high-precision measurements of the baryon density, a model of BBN must be assumed. However, beyond-standard models involving, for example, additional relativistic particle species or dark matter particles, often imply different conditions in the two epochs that can lead to different \DtH{} expectation values \citep{Steigman:2013}. Therefore, studying both epochs observationally and comparing the independent constraints on the baryon density, is an important opportunity to discover or rule out physics beyond the Standard Model.

\section*{Acknowledgments}
We would like to thank N. Crighton for useful discussions and comments, and an anonymous referee for constructive comments.
This research is based on observations collected at the European Organisation for Astronomical Research in the Southern Hemisphere, Chile, proposal ID 077.A-0166 obtained by PIs Carswell, Kim, Haehnelt and Zaroubi. It is also based on observations collected with the Keck Observatory Archive (KOA), which is operated by the W. M. Keck Observatory and the NASA Exoplanet Science Institute (NExScI), under contract with the National Aeronautics and Space Administration. The Keck data was obtained by PIs Songaila, Cowie, Crighton and Tytler.  
MTM thanks the Australian Research Council for \textsl{Discovery Project} grant DP130100568 which supported this work. Parts of this research were conducted by the Australian Research Council Centre of Excellence for All-sky Astrophysics (CAASTRO), through project number CE110001020.




\bibliographystyle{mnras}
\bibliography{bibliography} 


\appendix
\section{Model} \label{app:Model}
\label{tab:model}

\begin{table*}
\centering
\begin{minipage}{140mm}
\caption{The best fit parameter values for the four component model.}\label{tab:allmodel}
\begin{tabular}{l l l r r r }
\hline
Component		& Redshift 				& Specie		& $\log(N)$\footnotemark[1]			& $b_\mathrm{turb}$ or $b_\mathrm{tot}$ $[\kms]$ 	& $T$ [$10^{4} ~\mathrm{K}$]\\ \hline
A				& $3.572135\pm0.000007$	& \Hi			& $17.575$ 			& $2.36\pm1.43$				& $1.64\pm0.03$\\
				&						& \Di			& $12.994$			& $2.36\pm1.43$				& $1.64\pm0.03$\\
				&						& \Cii		& $12.360$			& $2.36\pm1.43$				& $1.64\pm0.03$\\
				&						& \Siii		& $11.197$			& $2.36\pm1.43$				& $1.64\pm0.03$\\
				&						& \Feiii		& $12.502$			& $2.36\pm1.43$				& $1.64\pm0.03$\\
				&						& \Siiv		& $11.699$			& $2.36\pm1.43$				& $1.64\pm0.03$\\
				
B 				& $3.572268\pm0.000002$	& \Hi			& $17.311$			& $2.15\pm0.76$				& $1.81\pm0.42$ \\
				&						& \Di			& $12.730$			& $2.15\pm0.76$				& $1.81\pm0.42$ \\
				&						& \Cii		& $12.777$			& $2.15\pm0.76$				& $1.81\pm0.42$ \\
				&						& \Siii		& $12.031$			& $2.15\pm0.76$				& $1.81\pm0.42$ \\
				&						& \Feiii		& $12.810$			& $2.15\pm0.76$				& $1.81\pm0.42$ \\
				&						& \Siiv		& $12.678$			& $2.15\pm0.76$				& $1.81\pm0.42$ \\

C				& $3.572451\pm0.000002$ 	& \Hi			& $17.402$			& $4.51\pm0.23$				& $1.87\pm0.07$ \\
				&						& \Di			& $12.821$			& $4.51\pm0.23$				& $1.87\pm0.07$ \\
				&						& \Cii		& $13.160$			& $4.51\pm0.23$				& $1.87\pm0.07$ \\
				&						& \Siii		& $12.311$			& $4.51\pm0.23$				& $1.87\pm0.07$ \\								
				&						& \Feiii		& $13.220$			& $4.51\pm0.23$				& $1.87\pm0.07$ \\
				&						& \Siiv		& $12.822$			& $4.51\pm0.23$				& $1.87\pm0.07$ \\

D				& $3.572682\pm0.000020$	& \Hi			& $15.943$			& $1.58\pm8.66$				& $4.35\pm0.94$ \\
				&						& \Di			& $11.362$			& $1.58\pm8.66$				& $4.35\pm0.94$ \\
				&						& \Siii		& $11.242$			& $1.58\pm8.66$				& $4.35\pm0.94$ \\
				&						& \Siiv		& $11.082$			& $1.58\pm8.66$				& $4.35\pm0.94$ \\

\hline

Summed 			&						& Specie		& $\sum\log(N)$ \\ \hline
				&						& \Hi			& $17.925\pm0.006$ \\
				&						& \Di			& $13.345\pm0.006$ \\
				&						& \Cii		& $13.357\pm0.029$ \\
				&						& \Siii		& $13.345\pm0.006$ \\
				&						& \Feiii		& $13.419\pm0.029$ \\
				&						& \Siiv		& $13.080\pm0.005$ \\
				&						& \Oi		& $12.060\pm0.127$\footnotemark[2] \\
\hline

Velocity shift		&& Data\footnotemark[3]		& Shift [$\kms$] \\ \hline
				&& setup 2		& --- \\
				&& setup 3		& $-1.326 \pm 0.089$\\
				&& setup 5		& $0.714 \pm 0.066$ \\
				&& setup 10 		& $-0.491\pm0.064$ \\
				&& VLT			& $0.504\pm0.061$ \\
				&& VLT Lyman 5	& $1.122\pm0.138$ \\

\end{tabular}
\vspace{1cm}

\footnotemark[1]{For the summed column densities the individual uncertainties are not determined.}\\
\footnotemark[2]{Based on a constant ratio of \Oi/\Hi{} across all components.}\\
\footnotemark[3]{setup 2, setup 3, setup 5 and setup 10 corresponds to the four different stacks of the Keck observations in \tabref{data}, while all the VLT exposures are combined into one spectrum}

\end{minipage}
\end{table*}

\begin{table*}
\centering
\begin{minipage}{140mm}
\contcaption{The model}
\begin{tabular}{l l l l r r }

\hline
Blends 			& Redshift 	& Specie		& $\log(N)$	& b $[\kms]$  \\ \hline

\Siii{} 1193 \AA		& 3.48711   	& \Hi			& 13.048		& 36.93	\\
				& 3.48789   	& \Hi			& 12.073		& 9.77	\\
		
\Siii{} 1304 \AA		& 3.41150    	& \Hi			& 13.062		& 21.61	 \\

\Feiii{} 1122 \AA	& 3.22110     	& \Hi			& 14.952		& 21.04	 \\
				& 3.22283   	& \Hi			& 12.096		& 22.28	 \\
				& 3.22376    	& \Hi			& 14.191		& 20.91	 \\
		
\Siiv{} 1393 \AA		& 4.24275		& unknown	& 12.214		& 17.29	\\

												
\Hi{} 1216 \AA\ (Lyman $\alpha$)	& 3.58017		& \Hi			& 12.372		&  13.77 \\
							& 3.57973		& \Hi			& 12.571		&  30.64 \\
							& 3.57847		& \Hi			& 13.661		&  21.94 \\
							& 3.57775		& \Hi			& 13.476		&  17.68 \\
							& 3.57733		& \Hi			& 12.935		&  \footnotemark[1]21.91\\
							& 3.57668		& \Hi			& 13.644		&  20.45 \\
							& 3.57560		& \Hi			& 12.827		&  29.26 \\
							& 3.57503		& \Hi			& 12.270		&  15.70 \\
							& 3.57462		& \Hi			& 12.372		& \footnotemark[1]29.61 \\
							& 3.57336		& \Hi			& 13.465 		&  6.89  \\
							& 3.57308		& \Hi			& 14.999		&  \footnotemark[1]29.58\\
							& 3.56951		& \Hi			& 13.085		&  34.30 \\
							& 3.56935  	& \Hi			& 11.473		&   8.32 \\
							& 3.56825		& \Hi			& 12.331		&  27.72 \\
							& 3.56715		& \Hi			& 13.450		&  31.23 \\
							& 3.56639		& \Hi			& 14.084		&  29.03  \\
							& 3.56497		& \Hi			& 14.131		&  27.00 \\
							& 3.56471 	& \Hi			& 14.029		&  40.28 \\


\Hi{}  \AA\ (Lyman $\gamma$)		& 2.65662		& \Hi 		& 11.920		& 8.50 \\
							& 2.65613 	& \Hi 		& 13.487		& 26.67 \\
							& 2.65693		& \Hi 		& 12.459		& 14.64  \\
							& 2.65921 	& \Hi 		& 12.526		&  8.41  \\
							& 2.65974		& \Hi 		& 14.345		& 37.06  \\
							& 2.66065		& \Hi 		& 12.686 		& \footnotemark[1]20.00   \\
							& 2.66101		& \Hi 		& 12.841		& 9.13    \\
							& 2.66139 	& \Hi 		& 12.579		& 12.88 \\

\Hi{}  \AA\ (Lyman 4)				& 2.56977  	& \Hi 		& 13.442		& 27.41 \\
							& 2.57124		& \Hi 		& 13.847 		& 23.74  \\
							& 2.57307		& \Hi 		& 13.026		& \footnotemark[1]20.00 \\
							& 2.57378  	& \Hi 		& 14.632		& 30.51    \\
							& 2.57449 	& \Hi 		& 13.727  		& 23.11  \\

 \Hi{}  \AA\ (Lyman 5)			& 2.52487		& \Hi 		& 13.374		& 26.72   \\
							& 2.52528 	& \Hi 		& 13.154		& 22.48   \\
							& 2.52577   	& \Hi 		& 12.834 		& 12.51  \\
							& 2.52609 	& \Hi 		& 12.916		& 20.95  \\
							& 2.52794  	& \Hi 		& 12.845		& \footnotemark[1]20.21\\
	
 \Hi{}  \AA\ (Lyman 6)			& 2.49914		& \Hi 		& 13.443		& 58.82   \\
							& 2.49945		& \Hi 		& 13.548  		& 23.38 \\
							& 2.50011		& \Hi 		& 11.793 		& 1.31 \\
							& 2.50224 	& \Hi 		& 13.156		& 26.30   \\
							& 2.50269 	& \Hi 		& 12.542		& 17.98  \\

\Hi{}  \AA\ (Lyman 7)				& 2.48172 	& \Hi 		& 11.550		& 1.00   \\
							& 2.48219		& \Hi 		& 12.755		& 27.04  \\
							& 2.48305		& \Hi 		& 12.312		& 12.89  \\
							& 2.48432		& \Hi 		& 13.050  		& \footnotemark[1]20.00  \\
							& 2.48512 	& \Hi 		& 14.446 		& 31.57   \\

\Hi{}  \AA\ (Lyman 8)				& 2.46946   	& \Hi 		& 13.255 		& 26.35 \\
							& 2.47150 	& \Hi 		& 13.541	 	& 20.62  \\
							& 2.47101		& \Hi 		& 13.213		& 31.34 \\
							& 2.47292		& \Hi 		& 13.177		& 31.21  \\
							& 2.47437		& \Hi 		& 13.019		& \footnotemark[1]40.65 \\
							& 2.47492		& \Hi 		& 13.812   	& 30.13  \\
							& 2.47534 	& \Hi 		& 13.489		& 19.67  \\
							
\Hi{}  \AA\ (Lyman 9)				& 2.46214 	& \Hi 		& 12.992	 	& 39.16 \\
							& 2.46265 	& \Hi 		& 12.744	& 16.01  \\
							& 2.46317		& \Hi 		& 14.130 	& 29.01 \\
\hline
\end{tabular}

\footnotemark[1]{A few of the $b$-parameters were fixed to prevent them from getting unphysically large.}\\
\end{minipage}
\end{table*}


\bsp	
\label{lastpage}
\end{document}